\documentclass[aps,twocolumn,superscriptaddress,floats,prd,nofootinbib,showpacs]{revtex4-2}

\usepackage{anyfontsize} 

\usepackage[utf8]{inputenc}
\usepackage[T1]{fontenc}
\usepackage{amssymb}
\usepackage{stmaryrd}
\usepackage{amsmath}
\usepackage{amsfonts}
\usepackage{mathrsfs}
\usepackage{amsmath,amssymb,amsfonts}
\usepackage{graphicx}
\usepackage{subfigure}
\usepackage{color} 
\usepackage{fancyhdr}
\usepackage{hyperref} 
\usepackage{booktabs}   
\usepackage{multirow}
\begin{document}

	\title{Quasinormal modes and tidal Love numbers of covariant effective quantum black holes with cosmological constant}
	
	\author{Yunlong Liu}
	\affiliation{School of Physics and Optoelectronics, South China University of Technology, Guangzhou 510641, China}
	\author{ Xiangdong Zhang}
	\email[Corresponding author.]{scxdzhang@scut.edu.cn}
	\affiliation{School of Physics and Optoelectronics, South China University of Technology, Guangzhou 510641, China}

	
	\begin{abstract}
		This paper systematically investigates the quasinormal modes (QNMs) and tidal Love numbers of covariant quantum-modified black holes (BHs) within two cosmological constant-dependent metric frameworks. By deriving axial/polar perturbation equations and employing pseudospectral methods (for QNMs) and numerical integration (for Love numbers), we quantify the quantum parameter $\zeta$'s influence. We find that, higher overtones exhibit enhanced $\zeta$ sensitivity. Solution 1 maintains purely imaginary QNMs under $\zeta$ variation, whereas Solution 2 demonstrates mode transitions from real to complex for polar/axial perturbations.
		Furthermore, for Love number, axial perturbations show non-monotonic $\zeta$ dependence, with $\Delta\mathcal{T}^a_l$ peaking at specific $\ell$ values before decay. While for polar perturbation of the same metric displays strictly monotonic $\zeta$ effects on $\Delta\mathcal{T}^p_l$.
These results highlight quantum gravity's role in BH perturbation dynamics and cosmological BH phenomenology.
	\end{abstract}
	
	\maketitle

	\section{Introduction}\label{Intro}
	Einstein's General Relativity (GR) faces fundamental limitations as a theory of spacetime, for instance, matter is described by quantum theory while spacetime remains classical, lacking a unified framework. Moreover, GR also suffers singularity crisis: GR predicts gravitational singularities where physical laws break down. These challenges motivate Loop Quantum Gravity (LQG) as a promising quantum gravity candidate, with its background independence and non-perturbative nature\cite{Background_Ashtekar_2004,Quantum_Rovelli_2004,Modern_Thiemann_2007,Fundamental_Han_2007}. During the last decades, application of methods of LQG to black hole models becomes a prosperity field and many black hole models with loop quantum corrections have been proposed \cite{Black_Gambini_2008,Quantum_Ashtekar_2018,Loop_Zhang_2020,Effective_Kelly_2020,Loop_Zhang_2023,Effective_Lin_2024}. However, most LQG black hole models suffer from implicit covariance issues \cite{Black_Gambini_2008,Quantum_Ashtekar_2018,Loop_Zhang_2020,Effective_Kelly_2020,Loop_Zhang_2023,Effective_Lin_2024}, where physical predictions depend on coordinate choices. Recent advances in Refs. \cite{Black_Zhang_2025,Black_Zhang_2025a,Black_Belfaqih_2025} have developed covariant black hole solutions under Hamiltonian constraints, later extended to include cosmological constants \cite{Effective_Lin_2024,Black_Zhang_2025a,Mass_Lin_2025,Covariant_Yang_2025}.

	In another respect, realistic black holes exist in dynamic environments. Specifically, dust clouds may surround these black holes, or they may lie in a larger-scale tidal field.
	This means that realistic black holes are perpetually in a perturbed state and engender various significant physical phenomena.
	
	These responses can be divided into two categories: one is quasi-normal modes (QNMs) associated with dynamic gravitational wave (GW), and the other is tidal Love numbers (TLNs) associated with static black hole deformations.
	
	On one hand, QNMs are a set of complex eigenvalues encoding gravitational wave (GW) frequencies (real part) and decay rates (imaginary part). They are crucial for probing the stability and structure of black holes. Investigations into QNMs hold substantial significance and interest across a range of fields.
	For example, through the detection of QNMs via gravitational waves, the measurement of black hole parameters and the testing of GR can be achieved \cite{Gravitational-wave_Echeverria_1989,Gravitational-wave_Berti_2006,Matched_Berti_2007,GW250114_Abac_2025}. 
	Moreover,  the search for QNMs in analogue gravity experiments (and other analogous systems) provides an experimental reference for the study of black hole QNMs, such as black hole spectroscopy in GW detection \cite{Quasinormal_Torres_2020}.

	On the other hand, TLNs, which reveal the static deformations under external tidal fields, originally defined in Newtonian gravity as dimensionless parameters quantifying an object's response to external gravitational perturbations \cite{yielding_Love_1909}, have been generalized to the framework in GR \cite{Tidal_Fang_2005,Relativistic_Binnington_2009,Relativistic_Damour_2009}. Interestingly, in GR, asymptotically flat black holes exhibit vanishing tidal Love numbers (TLNs) as a consequence of hidden symmetries in the spacetime geometry. This remarkable property stems from the unique mathematical structure of the Kerr solution, where the interplay between the no-hair theorem and the special algebraic properties of the metric enforces a perfect rigidity against static tidal deformation \cite{Hidden_Charalambous_2021,Love_Charalambous_2022,Ladder_Rai_2024,Exploring_Sharma_2024}. 
	However, in higher dimensions \cite{Black_Kol_2012,Static_Hui_2021,Scalar_Charalambous_2023,Love_Charalambous_2024,Magic_Charalambous_2024,Charging_Ma_2025}, modified gravity theories\cite{Testing_Cardoso_2017,Black_Cardoso_2018,Implications_DeLuca_2023,Tidal_Barura_2024}, the presence of a cosmological constant \cite{Asymptotically_Nair_2024,Tidal_Franzin_2024}, and celestial bodies surrounded by matter \cite{Probing_Baumann_2019,Environmental_Cardoso_2020,Perturbations_Capuano_2024,Flea_DeLuca_2024}, the hidden symmetry is broken, and hence non-zero TLNs could emerge. 
	
	Emerging theoretical and observational evidence indicates that quantum gravity effects significantly alter black hole perturbation responses \cite{Strong_Paul_2025,Light_Liu_2025,Long-lived_Lutfuoglu_2025,Mass_Lin_2025,Transition_Konoplya_2025,Loop_Motaharfar_2025}. Future gravitational wave experiments probing quasinormal modes (QNMs) and tidal Love numbers (TLNs) may provide direct signatures of quantum imprints in the black hole exterior. In this work, we systematically investigate the QNMs and TLNs of covariant quantum-modified black hole models with cosmological constant. Our study aims to quantify the influence of loop quantum gravity parameter $\zeta$ on perturbation spectra characterize its dependence on multipole moments $\ell$ and overtone indices n and establish phenomenological constraints for quantum-corrected black hole models.

	The paper is structured as follows:
	In Sec. \ref{SandW}, we give the covariant LQG metric and obtain the  perturbation equations.
	In Sec. \ref{QNMsR} and \ref{TLNsR}. we calculate the QNMs through the pseudospectral method and the TLNs with numerical method. 
	We summarize our results and draw
	concluding remarks in Sec. \ref{conclusion}.
	Throughout this work, we adopt natural units $c=G=1$.
	
	\section{Covariant effective spacetimes}\label{SandW}
	
	\subsection{Covariant effective spacetimes}
	To address the covariance challenge in effective quantum-corrected black hole construction, recent studies \cite{Black_Zhang_2025,Black_Zhang_2025a,Black_Belfaqih_2025} have developed a special Hamiltonian constraint framework yielding multiple spacetime solutions. This approach was further extended to the case with cosmological constant in \cite{Mass_Lin_2025,Covariant_Yang_2025} generating several covariant effective spacetime models. In the present work, we focus exclusively on the first two widely investigated cases from \cite{Black_Belfaqih_2025,Mass_Lin_2025,Covariant_Yang_2025}, which serve as tractable benchmarks for probing quantum gravitational effects. In the spherical coordinates $ ( t,r, \theta, \phi )$, the effective quantum corrected metric is given by
	\begin{align}
		ds^2 = - f[r] dt^2 + \frac{1}{f[r] h[r]}dr^2  + g[r]\left( d\theta ^2 + \sin^2\theta  d\phi ^2 \right), \label{CLQGbh}
	\end{align}
	where
	\begin{align}
	&\text{Solution 1:} \notag\\
		&f[r]=\left(1-\frac{2 M}{r}-\frac{\Lambda r^2}{3}\right)\left(1+ \frac{\zeta^{2}}{r^2} \left(1-\frac{2 M}{r}-\frac{\Lambda r^2}{3}\right) \right),  \notag\\
		&h[r]=1,  \quad g[r]=r^2;\\
	&\text{Solution 2:} \notag\\
		&f[r]=1-\frac{2 M}{r}-\frac{\Lambda r^2}{3}, \notag\\
		&h[r]=1+ \frac{\zeta^{2}}{r^2} \left(1-\frac{2 M}{r}-\frac{\Lambda r^2}{3}\right), \notag\\
		&g[r]=r^2.
	\end{align}
	Here, $M$ is the mass of the black hole, and $\zeta$ denotes the loop quantum parameter proportional to $\sqrt{\hbar}$. Solution 1 was first found in \cite{Mass_Lin_2025} and consistent with the results in Ref. \cite{Covariant_Yang_2025}. Meanwhile, solution 2 was discovered in Ref.\cite{Black_Belfaqih_2025} and  Ref.\cite{Covariant_Yang_2025}.
	In subsequent calculations, we use mass normalized parameters to reexpress $\hat{r}=r/M,\quad \hat{\zeta}=\zeta/M, \quad \hat{\Lambda}=M^2 \Lambda$, and ignore the overhat index $\hat{}$.   
	
	\subsection{Perturbation equations}
	The study of perturbations in black hole spacetime is roughly divided into two types:  
	
	1. Adding an external test field, and obtaining the perturbation equation through the generally covariant Klein-Gordon equation \cite{Classical_Landau_1971}, such as scalar and vector field perturbations. Here, ``test'' means that the backreaction of this external field on spacetime is negligible.

	2. Perturbing the background metric itself. 
	This is related to the oscillatory signals of gravitational waves emitted by the black hole. 
	In a spherically symmetric spacetime, perturbations can be expanded in terms of scalar, vector, and tensor spherical harmonics \cite{Stability_Regge_1957,Gravitational_Mathews_1962,Tensor_Zerilli_1970,Gravitational_Rezzolla_2003}.
	In the expansion process, tensor modes can be classified into two types by the polarity of the tensor functions: axial (odd) perturbations and polar (even) perturbations\cite{Gravitational_Zerilli_1970,Gravitational_Rezzolla_2003}.  
	
	Although the meanings of these two types of perturbations are different, in a spherically symmetric spacetime, they both ultimately lead to a similar second-order differential equation:  
	\begin{align} \label{timedomain}
		-\frac{\partial^{2} \Phi[t,r_*]}{\partial t^{2}}+\frac{\partial^{2} \Phi[t,r_*]}{\partial r_{*}^{2}}-V[r] \Phi[t,r_*]=0.
	\end{align}
	Further separating the time variable by assuming $\Phi[t,r_*]=e^{-i \omega t} \Psi[r_*]$, the above equation simplifies to 
	\begin{align} \label{MainEQ}
		\frac{\partial^{2} \Psi[r_*]}{\partial r_{*}^{2}}+\left(\omega^{2}-V[r]\right) \Psi[r_*]=0.
	\end{align}
	Here, $r_{*}$ is the tortoise coordinate, defined as $dr_{*}=dr/(f[r]\sqrt{h[r]})$.
	In the solution process, we assume that the boundary conditions are satisfied:
	\begin{align}\label{BC}
		\Psi[r_*] \sim e^{ \pm i \omega r_{*}}, \quad r_{*} \to \pm \infty .
	\end{align}
	This means that at the horizon, there are purely in-going waves, and at infinity, there are purely out-going waves.  
	
	The QNM frequency $\omega=\omega_{Re}-i \omega_{Im}$ represents the solution to the wave equation \eqref{MainEQ} satisfying the boundary conditions \eqref{BC}. This complex formulation naturally separates the oscillation and damping characteristics of gravitational waves. The real part $\omega_{Re}$ quantifies the actual oscillation frequency of the mode, governing the temporal evolution of the wave function. For instance, when computing gravitational wave frequencies from black hole perturbations, $\omega_{Re}$ directly corresponds to the wave's oscillation frequency. The imaginary part $\omega_{Im}$ encodes the mode's damping rate: a positive value signifies exponential decay of the wave function $\Psi$ with time, indicating perturbation dissipation and system stabilization, while a negative value reveals exponential growth, marking instabilities critical for black hole stability analyses.

	\subsubsection{Perturbation of external fields}

	The Klein-Gordon (KG) equation is used to describe perturbations \cite{Classical_Landau_1971}, which reads 
	\begin{align}
		\frac{1}{\sqrt{-g}} \partial_{\mu}\left(\sqrt{-g} g^{\mu \nu} \partial_{\nu} \psi\right)=0, \\
		\frac{1}{\sqrt{-g}}\partial_{\mu}(F^{\rho\sigma}g_{\rho\nu}g_{\sigma\mu}\sqrt{-g}) = 0,
	\end{align}
	where $\psi$ is a massless scalar field, and $F_{\mu\nu}=\partial_{\mu}A_{\nu}-\partial_{\nu}A_{\mu}$ is the electromagnetic tensor.
	By separating variables with spherical harmonics and separating the time variable, the equation can be reduced to the form of Eq.\eqref{MainEQ} as follows: 
	\begin{align}
		\label{ScaEq}
		\Psi_s''[r_*] = (V_{s}[r]-\omega^2)\Psi_s[r_*].
	\end{align}	 
	Here, The corresponding effective potential $V_{s}$ is in the following form:   
	\begin{widetext}
	\begin{align}
		\label{Vs}
		V_s[r] =  f[r] \left(\frac{(\ell -1) (\ell +2)+2}{r^2} + \frac{1-s}{2 r} \left(2 h[r] f'[r]+f[r] h'[r]\right) \right),
	\end{align}
	\end{widetext}
	where  $s$ denotes the spin of the test field  and  $\ell$ represents the multipole index satisfying the condition $\ell \geq s$. When $s=0$, it represents the scalar field perturbation equation.   
	
	\subsubsection{axial and polar perturbations}

	In LQG black hole models, the effective metric is derived by an effective Hamiltonian constraints with loop quantum corrections. 
	Under certain circumstances, quantum corrections can be effectively interpreted as an anisotropic energy-momentum tensor fluid within the framework of Einstein's gravitational theory \cite{Noncommutative_Nicolini_2006, Quantum_Ashtekar_2018, Quantum_Ashtekar_2018a, Gravitational_Chen_2019}. 
	That is, taking the solution \eqref{CLQGbh} as the solution to Einstein's field equations, as follows:
	\begin{align}
		G_{\mu\nu}=R_{\mu\nu}-\frac{1}{2}g_{\mu\nu}R + 
		g_{\mu\nu} \bar{\Lambda}= 8\pi T_{\mu\nu}^{eff}.
		\label{ET}
	\end{align}
	Here, $\bar{\Lambda}$ is the effective cosmological constant.   
	In this paper, we assume that it is related to the properties of $f[r]$ at infinity. 
	For example, for solution 1, the expansion at infinity is $\Lambda(1-{\zeta^2 \Lambda }/{3})r^2+...$, which means the effective cosmological constant is $\bar{\Lambda}=\Lambda(1-{\zeta^2 \Lambda }/{3})$.
	Along the same line, in solution 2, the effective cosmological constant is just $\bar{\Lambda}=\Lambda$. Note that there is ambiguity in the definition of $\bar{\Lambda}$ in solution 2, which is because $h[r] \neq 1$ in solution 2. This makes it necessary to be very careful when calculating the TLNs of solution 2.
	
	In addition, the corrections of LQG can be equivalent to the anisotropic fluid $T_{\mu \nu}^{eff}$.
	For example, in solution 1,
	\begin{align}
		T_{\mu \nu}^{eff}=\text{diag}\left[f[r]\,T_{a}[r],\,-\frac{T_{a}[r]}{f[r]},\,T_{b}[r],\,T_{b}[r]\sin^2[\theta] \right],
	\end{align}
	where
	\begin{align}
		T_{a}[r] &= \frac{\zeta^2 \left( 36  - 24  r + 3 r^2 + 2 \Lambda r^4 \right)}{3 r^6}, \\
		T_{b}[r] &= \frac{\zeta^2 \left( 24  - 12  r + r^2 \right)}{r^4}.
	\end{align}
	When the loop quantum parameter $\zeta$ vanishes, $T_{\mu \nu}^{eff}=0$, and we can recover the classical vacuum Einstein equation with a cosmological constant.  
	This equivalence makes the study of effective gravitational perturbations possible \cite{Gravitational_Cruz_2019,Polar_Cruz_2020,Breaking_del-Corral_2022}. 
	By assuming that the perturbation of the anisotropic fluid is negligible in the gravitational perturbation, the metric is then perturbed and analyzed as follows. 
	
	First, the new spacetime metric $\tilde{g}_{\mu \nu}$ is written as the sum of the original background metric $g_{\mu \nu}$ and the perturbation part $h_{\mu \nu}$, as follows:
	\begin{align}\label{dg}
		\tilde{g}_{\mu \nu}=g_{\mu \nu}+h_{\mu \nu},
	\end{align}
	where the perturbed metric $h_{\mu \nu}$ satisfies $|h_{\mu \nu}| \ll |g_{\mu \nu}|$.
	
	Considering the spherical symmetry of the background metric, the perturbation $h_{\mu \nu}$ can be decomposed into two independent parts: axial perturbation $h_{\mu \nu}^{axial}$ and polar perturbation $h_{\mu \nu}^{polar}$.
	In the Regge–Wheeler gauge \cite{Stability_Regge_1957,Gravitational_Zerilli_1970,Gravitational_Rezzolla_2003}, these two parts can be written as:
	
	\begin{widetext}
	\begin{align}
		h_{\mu \nu}^{pol}&=
		\begin{pmatrix}
			 H^p_{0}[r] f[r] & H^p_{1}[r] & 0 & 0 \\ 
			H^p_{1}[r] & {H^p_{2}[r]}/{(f[r]h[r])}  & 0 & 0 \\ 
			0 & 0 & K[r] g[r] & 0 \\ 
			0 & 0 & 0 & K[r] g[r] \sin^{2}[\theta]
		\end{pmatrix}
		P_{l}[\cos[\theta]] e^{-i \omega t}, \label{hpol}\\
		h_{\mu \nu}^{axial }&=
		\begin{pmatrix}
			0 & 0 & 0 & H^a_{0}[r] \\
			0 & 0 & 0 & H^a_{1}[r] \\
			0 & 0 & 0 & 0 \\
			H^a_{0}[r] & H^a_{1}[r] & 0 & 0
		\end{pmatrix}
		 \partial_{\theta} P_{l}[\cos[\theta]] \sin[\theta]   e^{-i \omega t} . \label{haxial}
	\end{align}
	\end{widetext}
	
	First, assuming $H^p_{0}[r]=H^p_{2}[r]$ and substituting equations \eqref{hpol} and \eqref{dg} into equation \eqref{ET}, we can obtain four equations containing the independent variables $\{H^p_0[r],H^p_1[r],K[r],H'^p_0[r],H'^p_1[r],K'[r]\}$,
	by redefining the functions \cite{Polar_Cruz_2020}, as  
	\begin{align}
		K[r]=F_{KK}[r] \bar{K}[r_*]+ F_{KR}[r]  \bar{R}[r_*], \\
		H^p_1[r]/\omega=R[r]=F_{RK}[r] \bar{K}[r_*]+ F_{RR}[r]  \bar{R}[r_*],
	\end{align}
	where 
	\begin{widetext}
		\begin{align}
			F_{KK}[r] &= -\sqrt{f[r]^{2}h[r]} \left( F_{A0}[r]+\frac{F_{B0}[r]F_{A2}[r]}{F_{B2}[r]} \right), \quad F_{KR}[r] = 1, \\
			F_{RK}[r] &=  
			\frac{1}{F_{B2}[r]} \left( \frac{1}{\sqrt{f[r]^2 h[r]}} + \sqrt{f[r]^2 h[r]} F_{A2}[r] \left( F_{A0}[r] -  F_{B0}[r] \frac{F_{A2}[r] }{F_{B2}[r]} \right) \right), \quad 	F_{RR}[r] =  - \frac{F_{A2}[r]}{F_{B2}[r]}.
		\end{align}
	\end{widetext}
	In the above equations, the unknown functions are still $F_{A0}[r]$, $F_{B0}[r]$, $F_{A2}[r]$, and $F_{B2}[r]$.
	Their specific expressions are as follows: 
	\begin{widetext}
		\begin{align}
			F_{A0}[r]&= \frac{1}{f[r] g[r] F_f[r]} \Big( -2 \bar{\Lambda} g[r]^2 f'[r] + f[r] h[r] f'[r] g'[r]^2 \notag\\
			&- g[r] \left( (-2 + \ell + \ell^2) f'[r] - 2 \bar{\Lambda} f[r] g'[r] + h[r] f'[r]^2 g'[r] \right) \Big)  ,\\
			F_{B0}[r]&=- \frac{1}{2 g[r]^2 F_f[r]} \mathrm{i} \left( -2 + \ell + \ell^2 + 2 \bar{\Lambda} g[r] + h[r] f'[r] g'[r] \right) \Big( 4 \bar{\Lambda} g[r]^2 - f[r] h[r] g'[r]^2 \notag\\
			& + 2 g[r] \left( -2 + \ell + \ell^2 + f[r] g'[r] h'[r] + h[r] \left( f'[r] g'[r] + 2 f[r] g''[r] \right) \right) \Big) ,\\
			F_{A2}[r]&= \frac{2 g'[r]}{f[r] F_f[r]}; \quad F_{B2}[r]=- \frac{\mathrm{i} h[r] g'[r]^2}{g[r] F_f[r]}  ,\\
			F_f[r]&=3 h[r] f'[r] g'[r]+4 \bar{\Lambda}  g[r]+2 \left(\ell ^2+\ell -2\right).
		\end{align}
	\end{widetext}
	 
	Next, we add an additional assumption,
	\begin{align}
		\bar{K}'[r_*]&=\bar{R}[r_*],\\
		\bar{R}'[r_*]&=(V_{p}-\omega^2)\bar{K}[r_*].
	\end{align}	
	Let $\Psi_p=\bar{K}$, and we can obtain   
	\begin{align}\label{PloEq}
		\Psi_p''[r_*] = (V_{p}[r]-\omega^2)\Psi_p[r_*].
	\end{align}	
	where
	\begin{widetext}
		\begin{align}\label{Vp}
			V_p[r]=
			-\sqrt{f[r]^{2}h[r]}\left(\frac{F_{B0}[r]}{\sqrt{f[r]^{2}h[r]}F_{B2}[r]}+\left(F_{A0}[r]+\frac{F_{A2}[r]F_{B0}[r]}{F_{B2}[r]}\right)F_{KK}[r]+F'_{KK}[r]\right).
		\end{align}
	\end{widetext}
	Here $\ell$ is the angular quantum number, $\ell \geq 2$.   
	When the effective cosmological constant $\bar{\Lambda}$ tends to zero, the results in reference \cite{Polar_Cruz_2020} can be recovered.

	Similarly, substituting equations \eqref{haxial} and \eqref{dg} into equation \eqref{ET}, we can obtain two equations containing the independent variables $\{H^a_0[r],H^a_1[r],H'^a_0[r],H'^a_1[r]\}$.  
	Eliminating the variables $\{H^a_0[r],H'^a_0[r]\}$ from the equations and redefining the $H^a_1[r]$ as
	\begin{align}
		\Psi_a[r_*] = {f[r] \sqrt{h[r]}} H^a_1[r] /{\sqrt{g[r]}}.
	\end{align}
	We can obtain
	\begin{align}\label{AxiEq}
		\Psi_a''[r_*] = (V_{a}[r]-\omega^2)\Psi_a[r_*].
	\end{align}	
	Here, the effective potential $V_{a}[r]$ can be described as
	\begin{widetext}
		\begin{align}\label{Va}
			V_a[r]&=
			\frac{1}{4g[r]^2}\bigg(2f[r] g[r]\Big(2(-2 + \ell+\ell^2)+h[r]f'[r]g'[r]+g[r]\big(4\bar{\Lambda} + f'[r]h'[r]+2h[r]f''[r]\big)\Big)\notag\\
			&+f[r]^2\big(g[r]g'[r]h'[r]+h[r]\big(g'[r]^2 + 2g[r]g''[r]\big)\big)\bigg).
		\end{align}
	\end{widetext}
	When the effective cosmological constant $\bar{\Lambda}$ tends to zero, the results in reference \cite{Gravitational_Cruz_2019} can be recovered.
	
	\section{Quasinormal modes}\label{QNMsR}
	 
	In this section, we will mainly focus on the QNMs of the covariant LQG metric \eqref{CLQGbh} with a cosmological constant.  
	The calculation of  QNMs has been approached through diverse schemes, with origins dating back to the early 1970s.  
	Blome and Mashhoon approximated the black hole potential with a known P\"oschl-Teller potential \cite{Quasinormal_Blome_1984}.  
	Soon,  QNMs were obtained via the WKB approximation method \cite{Black_Schutz_1985}, with subsequent studies \cite{Blackhole_Iyer_1987,Quasinormal_Konoplya_2003,Quasinormal_Matyjasek_2017,Higher_Konoplya_2019} proposing higher-order WKB approximation methods.   
	
	However, in the case of high overtone, the calculation error of the WKB method will increase, and a more reliable and rapid method is the Chebyshev pseudospectral method \cite{Overdamped_Jansen_2017}. 
	This scheme compresses the interval from the event horizon and the cosmological horizon to $(0,1)$ by introducing a new variable $z$ to replace $r$, and the detailed procedure is provided in Appendix \ref{CPS}.
	Then, the wave function is discretized on the Chebyshev-Lobatto grid points $z_j = (1 - \cos[\pi j/N])/2$.  
	The discrete matrix equation is obtained:
	\begin{align}
		\left(\mathcal{M}_{0} + \mathcal{M}_{1}\omega + \mathcal{M}_{2}\omega^{2}\right)y = 0,
	\end{align}
	where $y$ is the numerical vector of the unknown function, and $\mathcal{M}_{i}$ are the numerical matrices at the grid points.
	By directly solving this generalized eigenvalue equation, the quasinormal modes can be obtained.
	
	Other methods for calculating QNMs include the continued fraction method developed by Leaver \cite{analytic_Leaver_1985} and perfected by Nollert \cite{Quasinormal_Nollert_1993}.   
	The direct integration method proposed by Pani \cite{Advanced_Pani_2013}.
	More descriptions of calculation methods can be found in the more recent review articles \cite{Quasinormal_Berti_2009,Quasinormal_Konoplya_2011}.   
	
	In the present paper, we will mainly use the pseudospectral method \cite{Overdamped_Jansen_2017} for calculations.
	
	\subsection{Solution 1}
	
\begin{figure*}[!htb]
		\centering
		\subfigure[]{
			\includegraphics[width=0.31\textwidth]{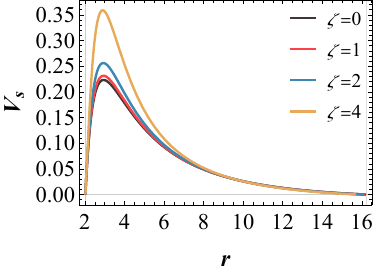}
		}
		\subfigure[]{
			\includegraphics[width=0.31\textwidth]{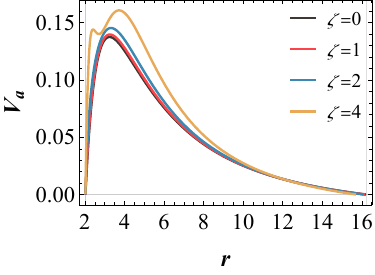}
		} 
		\subfigure[]{
			\includegraphics[width=0.31\textwidth]{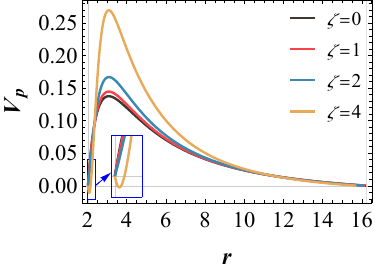}
		}
		\caption{Effective potential of  (a) scalar field, (b) axial, and (c) polar perturbations as a function of radius $r$ for $\zeta = \{0, 1, 2, 3\}$ with $\ell = 2$ and $\bar{\Lambda} = 0.01$.}
		\label{G1_Veff_r}
	\end{figure*}

	The effective potential can be shown in Fig.\ref{G1_Veff_r}.   
	It can be seen that the peaks of three effective potentials increase with the increase of $\zeta$.
	For axial perturbations, when the loop quantum parameter is large enough, the effective potential splits into two peaks. 
	For polar perturbations, when the loop quantum parameter is relatively large, a tiny potential well appears near the horizon.
	
\begin{figure*}[!htb]
		\centering
		\subfigure[]{
			\includegraphics[width=0.31\textwidth]{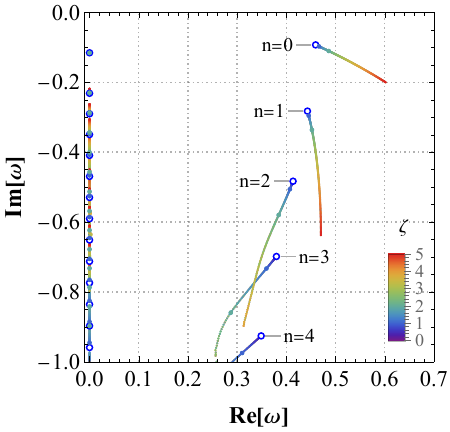}
			\label{G1_QNM_Vs_ReIm}
		}
		\subfigure[]{
		\includegraphics[width=0.31\textwidth]{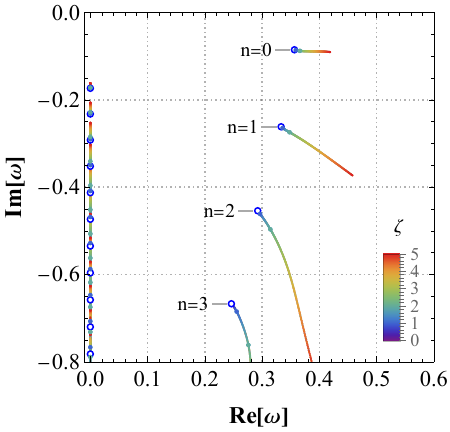}
		\label{G1_QNM_Va_ReIm}
		}
		\subfigure[]{
			\includegraphics[width=0.31\textwidth]{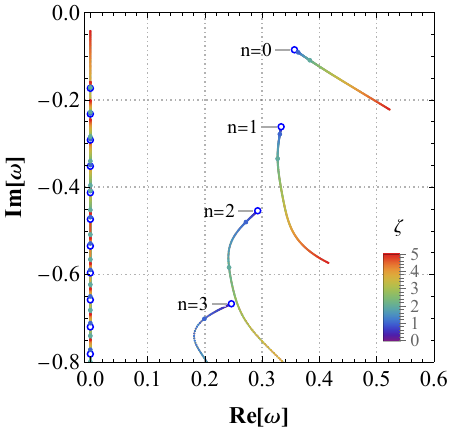}
			\label{G1_QNM_Vp_ReIm}
		}
		\caption{The QNMs trajectories of  (a) scalar field, (b) axial, and (c) polar perturbations in the complex $\omega$ plane for $\ell = 2$  and $\bar{\Lambda} = 0.01$. The hollow dot represent the QNM values  of dS black hole.}
	\end{figure*}

	In the complex $\omega$ plane Figs. \ref{G1_QNM_Vs_ReIm}-\ref{G1_QNM_Vp_ReIm}, we respectively show the quasinormal modes of scalar field perturbation, axial  and polar perturbations as the loop quantum parameter changes.
	From Fig. \ref{G1_QNM_Vs_ReIm}, we can see that for complex modes, the imaginary part decreases as  $\zeta$ increases, while the pure imaginary mode remains a pure imaginary mode as $\zeta$ increases.
	When the loop quantum parameter reaches a certain size, the pure imaginary mode becomes the dominant mode.  
	This is consistent with the conclusion in Ref. \cite{Mass_Lin_2025}.  
	Similarly, the same is true for polar perturbations in Fig. \ref{G1_QNM_Vp_ReIm}.   
	However, this situation does not occur in the axial case, because as $\zeta$ increases, the imaginary part of the fundamental mode decreases very slowly in Fig. \ref{G1_QNM_Va_ReIm}. 
	
	Furthermore, by comparing Figs. \ref{G1_QNM_Vs_ReIm} to \ref{G1_QNM_Vp_ReIm}, it can be found that the complex fundamental modes all vary linearly with the $\zeta$. However, when it comes to 
	$n=1$, the real parts of polar perturbations showing an non-monotonic behavior of decreasing first and then increasing. 
	This reflects that a stronger nonlinearity is implied under the complex expression of $V_p[r]$ in the polar perturbation equations.
	
	Finally, we give the specific values of the fundamental mode and the first few overtones in complex QNMs, as shown in Table \ref{G1_QNM_T}. 
	\begin{table*}[!htb]
		\centering
		\caption{\label{G1_QNM_T}
			The values of fundamental QNM and first three overtones for $\ell= 2$ and several values of $\zeta$  in solution 2.}
		\begin{ruledtabular}
			\begin{tabular}{lcccc} 
				$n$     &  $\zeta$  &  scalar   &   axial  &   polar\\
				\midrule 
			$0$	&	$1$	&	0.4664830 - 0.0972362i 	&	0.3588194-0.0863506i 	&	0.3630206-0.0914033i 	\\
				&	$2$	&	0.4863431-0.1103499i 	&	0.3659708-0.0879245i 	&	0.3832439-0.1094183i 	\\
				&	$3$	&	0.5172813-0.1320390i 	&	0.3786057-0.0887196i 	&	0.4180479-0.1388442i	\\
				&	$4$	&	0.5567902-0.1620596i 	&	0.3969233-0.0891740i 	&	0.4659422-0.1773590i 	\\
				\midrule 
			$1$	&	$1$	&	0.4455513 - 0.2957676i 	&	0.3357965-0.2647910i 	&	0.3309869-0.2787840i 	\\
				&	$2$	&	0.4521507 - 0.3363306i 	&	0.3479387-0.2744635i 	&	0.3271615-0.3349631i 	\\
				&	$3$	&	0.4602947 - 0.4057923i 	&	0.3740197-0.2952443i 	&	0.3356652-0.4211537i 	\\
				&	$4$	&	0.4673816 - 0.5060305i 	&	0.4114579-0.3289584i 	&	0.3555870-0.5064619i 	\\
				\midrule 
			$2$	&	$1$	&	0.4072963 - 0.5062308i 	&	0.2965141-0.4617311i 	&	0.2717109-0.4801554i 	\\
				&	$2$	&	0.3845442 - 0.5800146i 	&	0.3144783-0.4968371i 	&	0.2423007-0.5840104i 	\\
				&	$3$	&	0.3445477 - 0.7240719i 	&	0.3399823-0.5683484i 	&	0.2668001-0.6966517i 	\\
				&	$4$	&	0.3058163 - 0.9421864i	&	0.3639402-0.6785899i 	&	0.3672476-0.8404503i 	\\
			\end{tabular}
		\end{ruledtabular}
	\end{table*}
	
	\subsection{Solution 2}
	
	\begin{figure*}[!htb]
		\centering
		\subfigure[]{
			\includegraphics[width=0.31\textwidth]{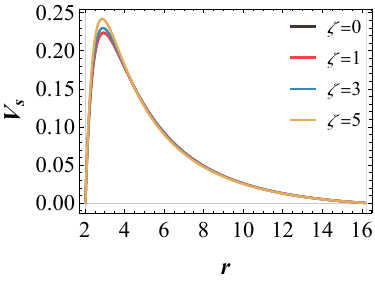}
		}
		\subfigure[]{
			\includegraphics[width=0.31\textwidth]{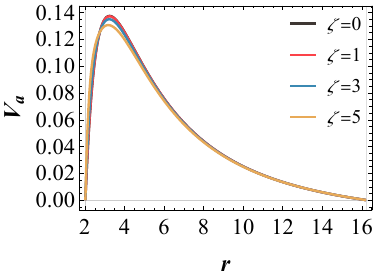}
		} 
		\subfigure[]{
			\includegraphics[width=0.31\textwidth]{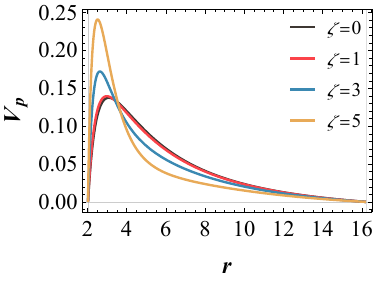}
		}
		\caption{Effective potential of  (a) scalar field, (b) axial, and (c) polar perturbations as a function of radius $r$ for $\zeta = \{0, 1, 3, 5\}$ with $\ell = 2$ and $\bar{\Lambda} = 0.01$.}
		\label{G2_Veff_r}
	\end{figure*}
	From the Fig. \ref{G2_Veff_r}, it can be seen that the peaks of the three effective potentials increase with the increase of $\zeta$. Unlike Solution 1, special structures like double peaks will not occur.
	
	\begin{figure*}[!htb]
		\centering
		\subfigure[]{
			\includegraphics[width=0.31\textwidth]{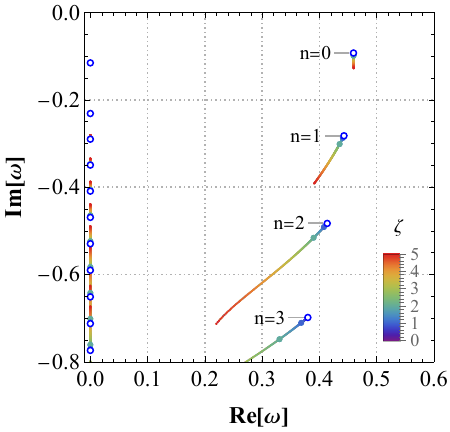}
			\label{G2_QNM_Vs_ReIm}
		}
		\subfigure[]{
			\includegraphics[width=0.31\textwidth]{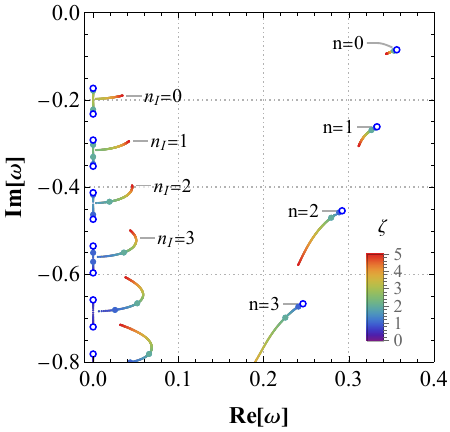}
			\label{G2_QNM_Va_ReIm}
		}
		\subfigure[]{
			\includegraphics[width=0.31\textwidth]{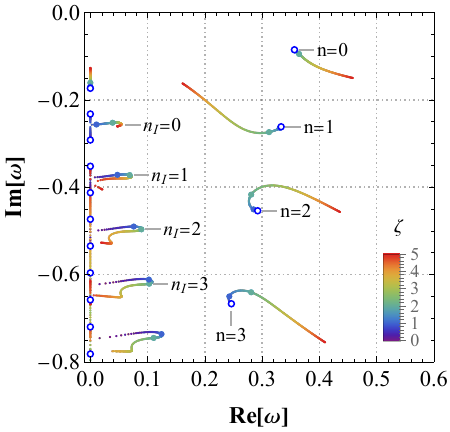}
			\label{G2_QNM_Vp_ReIm}
		}
		\caption{The QNMs trajectories of  (a) scalar field, (b) axial, and (c) polar perturbations in the complex $\omega$ plane for $\ell = 2$  and $\bar{\Lambda}  = 0.01$.
			The hollow dot represent the QNM values  of dS black hole.}
		\label{G2_QNM_ReIm}
	\end{figure*}
	Next, we consider the QNMs of solution 2. 
	In the complex $\omega$ plane Figs. \ref{G2_QNM_Vs_ReIm}-\ref{G2_QNM_Vp_ReIm}, we respectively show the quasinormal modes of scalar field , axial and polar perturbations as $\zeta$ changes.

	From Fig. \ref{G2_QNM_Vs_ReIm}, we can see that for complex fundamental modes, as $\zeta$ increases, the imaginary part decreases, while the real part shows a slight increase . For $n \geq 1$, the real part decreases significantly as $\zeta$ increases.
	
	Then, Fig. \ref{G2_QNM_Va_ReIm} shows that, as $\zeta$ changes, both the real and imaginary parts of $\omega$ for different $n$ values decrease monotonically. Meanwhile, the larger the $n$, the more significant the influence of $\zeta$. Note that the purely imaginary modes no longer remain purely imaginary; instead, they merge in pairs as $\zeta$ increases, eventually becoming complex QNMs, thereby forming a new $n_I$ branch.
	
\begin{figure*}[!htb]
		\centering
		\subfigure{
			\includegraphics[width=0.34\textwidth]{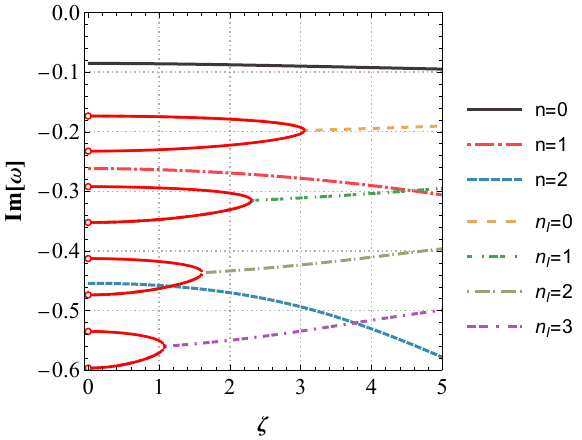}
			\label{G2_QNM_Va_Im}
		}
		\subfigure{
			\includegraphics[width=0.34\textwidth]{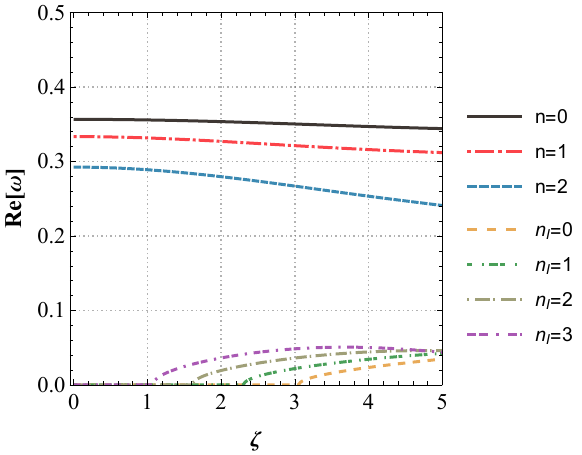}
			\label{G2_QNM_Va_Re}
		}
		\caption{The imaginary(left) and real(right) part of the QNMs as a function of $\zeta$ with $\ell = 2$  and $\bar{\Lambda}  = 0.01$ for axial perturbation.}
		\label{G2_QNM_Va_Im_Re}
	\end{figure*}

\begin{figure*}[!htb]
		\centering
		\subfigure{
			\includegraphics[width=0.34\textwidth]{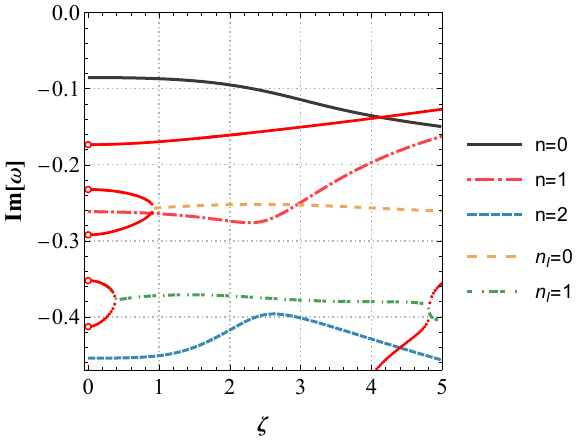}
			\label{G2_QNM_Vp_Im}
		}
		\subfigure{
			\includegraphics[width=0.34\textwidth]{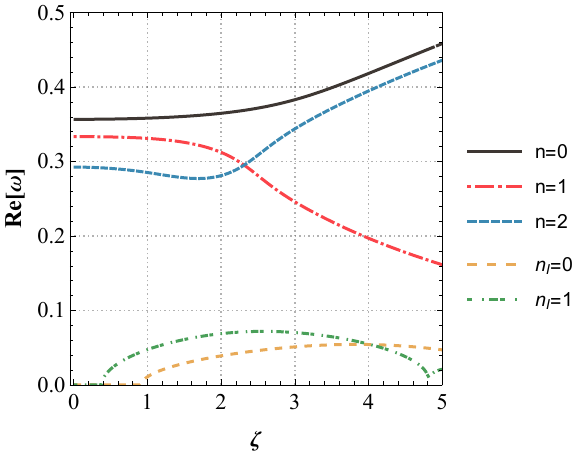}
			\label{G2_QNM_Vp_Re}
		}
		\caption{The imaginary(left) and real(right) part of the QNMs as a function of $\zeta$ for $\ell = 2$  and $\bar{\Lambda}  = 0.01$ for polar perturbation.}
		\label{G2_QNM_Vp_Im_Re}
	\end{figure*}

	In Fig. \ref{G2_QNM_Vp_ReIm}, the trajectory of $\omega$ as a function of $\zeta$ is more complex. On the one hand, similar to Solution 1, due to the complexity of the effective potential for polar perturbations, for $n \geq 1$, the trajectory of the complex $\omega$ no longer maintains monotonicity. Meanwhile, the purely imaginary modes also no longer remain purely imaginary and form a new  $n_I$ branch.
	More detailed results are presented in Figs. \ref{G2_QNM_Va_Im_Re}-\ref{G2_QNM_Vp_Im_Re}, showing that the imaginary and real parts of the QNMs as a function of $\zeta$ for the fundamental mode and different overtones.

	In the Figs. \ref{G2_QNM_Va_Im_Re}, the red solid lines in the left figure, which have no counterparts in the right one,  represent purely imaginary QNMs. It can be clearly seen that as $\zeta$ increases, the purely imaginary QNMs merge in pairs and form a new branch $n_I$.
	From the Figs.\ref{G2_QNM_Vp_Im_Re}, we can also find the newly emerged branch $n_I$ at the imaginary part of the QNMs. Unlike axial perturbations, when $\zeta$ is sufficiently large, the purely imaginary modes become the dominant modes. Whatsmore, we can observe that degeneracy occurs between $n=1$ and $n=2$ around $\zeta=2.3$ at the real part of the QNMs.
	
	Similarly, we also provide some numerical results of the complex modes of polar perturbations for $n=0,1,2,3$ and various values of $\zeta$, as shown in Table \ref{G2_QNM_T}.
	\begin{table*}[!htb]
		\centering
		\caption{\label{G2_QNM_T}
			The values of fundamental QNM and first three overtones for $\ell= 2$ and several values of $\zeta$  in solution 2.}
		\begin{ruledtabular}
			\begin{tabular}{lcccc} 
		$n$     &  $\zeta$  &  scalar   &   axial  &   polar \\
				\midrule 
			$0$	&	$1$	&	0.4596641 - 0.0944465i	&	0.3556679 - 0.0861155i 	&	0.3580325 - 0.0870656i 	\\
				&	$2$	&	0.4597267 - 0.0990951i 	&	0.3534051 - 0.0877773i 	&	0.3645969 - 0.0951733i 	\\
				&	$3$	&	0.4598201 - 0.1064708i 	&	0.3502179 - 0.0900195i 	&	0.3828083 - 0.1145923i 	\\
				&	$4$	&	0.4599333 - 0.1161632i 	&	0.3468506 - 0.0924648i 	&	0.4180797 - 0.1356564i 	\\
				\midrule 
			$1$	&	$1$	&	0.4411553 - 0.2872825i 	&	0.3314830 - 0.2637732i 	&	0.3309170 - 0.2644852i 	\\
				&	$2$	&	0.4354418 - 0.3014062i 	&	0.3268680 - 0.2693174i 	&	0.3121239 - 0.2741425i 	\\
				&	$3$	&	0.4255674 - 0.3240235i 	&	0.3211299 - 0.2781334i 	&	0.2457404 - 0.2497274i 	\\
				&	$4$	&	0.4110256 - 0.3542622i 	&	0.3158660 - 0.2903511i 	&	0.1968047 - 0.1972378i 	\\
				\midrule 
			$2$	&	$1$	&	0.4079979 - 0.4915302i 	&	0.2887864 - 0.4579567i 	&	0.2850852 - 0.4510489i 	\\
				&	$2$	&	0.3899642 - 0.5160741i 	&	0.2794774 - 0.4701175i 	&	0.2809232 - 0.4172991i 	\\
				&	$3$	&	0.3571364 - 0.5565598i 	&	0.2668220 - 0.4937008i 	&	0.3436366 - 0.4014712i 	\\
				&	$4$	&	0.3038188 - 0.6152810i 	&	0.2532794 - 0.5307060i 	&	0.3945858 - 0.4291363i 	\\
			\end{tabular}
		\end{ruledtabular}
	\end{table*}
	
	\section{Tidal Love numbers}\label{TLNsR}
	TLNs, which are used to quantify the deformability of self-gravitating objects, are defined as the linear response of the mass and current multipole moments of a body to an external tidal field. They are naturally extended to the framework of linear gravitational perturbation theory \cite{Tidal_Fang_2005,Constraining_Flanagan_2008,Relativistic_Binnington_2009,Relativistic_Damour_2009}.

	By taking the zero frequency limit, $\omega \rightarrow 0$, of equations \eqref{PloEq}-\eqref{AxiEq}, we can obtain the master perturbation equation that describes the small deformations of a physical system under the action of external forces (such as tidal fields).
	The solutions to this equation can be decomposed into two types of ``fundamental components'', namely non-normalizable modes and normalizable modes \cite{Tidal_Franzin_2024,Parametrized_Katagiri_2024,Loop_Motaharfar_2025}. 
	Non-normalizable modes, analogous to ``background driving terms'', are directly induced by external tidal fields and represent ``forced perturbations'' (such as the force you exert when pushing an object); in contrast, Normalizable modes, analogous to ``system response terms'', describe the ``rebound'' of the black hole itself to these forced perturbations (such as the vibration of an object after being pushed).
	Once these two types of modes are determined, any solution to the master perturbation equation can be expressed as their ``linear combination'' (i.e., a proportional mixture).
	
	TLNs are the dimensionless ratio of the ``normalizable mode coefficient'' (intensity of the response term) to the ``non-normalizable mode coefficient'' (intensity of the forced term) in the aforementioned linear combination. They are used to measure the ``sensitivity'' of a black hole to tidal fields — the larger the ratio, the more significant the deformation of the black hole under tidal action.
	
	In the following, note that the Love number of solution 1 and 2 without cosmological constant is obtained in \cite{Love_Motaharfar_2025} using perturbative approach and Green’s function technique, in order to make a comparison,  we will calculate the TLNs for solution 1 and 2 using the numerical method developed in Ref. \cite{Tidal_Franzin_2024} to perform the calculation of TLNs.
	
	\subsection{Results}
	
	For the AdS case, the asymptotic solutions of master perturbation equation at infinity can be expressed as 
	\begin{eqnarray}
		\Phi_a &=& D_a^+ \sum_{i=0}^{\infty} \frac{a^+_i}{r^i} + \frac{D_a^-}{r} \sum_{i=0}^{\infty} \frac{a^-_i}{r^i}, \label{PhiaInf}\\
		\Phi_p &=& D_p^+ \sum_{i=0}^{\infty} \frac{p^+_i}{r^i} + \frac{D_p^-}{r} \sum_{i=0}^{\infty} \frac{p^-_i}{r^i},	\label{PhipInf}
	\end{eqnarray}
	and the Taylor expansion at the horizon can be described as   
	\begin{eqnarray}
		\Phi_{a} &=&\sum_{i=0}^{\infty} a^h_{i}\left(\frac{r-r_{h}}{r_{h}}\right)^{i},	\label{Phiarh}\\
		\Phi_{p} &=&\sum_{i=0}^{\infty} p^h_{i}\left(\frac{r-r_{h}}{r_{h}}\right)^{i},	\label{Phiprh}
	\end{eqnarray}
	where $D_a^+$ and $D_p^+$ is the coefficient of the non-normalizable term (corresponding to the constant term in the asymptotic expansion that does not decay with  $r$), while $D_a^-$ and $D_p^-$  is the coefficient of the normalizable term (corresponding to the term in the asymptotic expansion that decays with  $1/r$) .
	The specific coefficient expressions of soluton 1 and 2 can be found in Appendix \ref{CTLN}.
	Corresponding to the Ref.\cite{Tidal_Franzin_2024}, we can define the axial and polar TLNs as
	\begin{align}
		\mathcal{T}^a_l = \frac{1}{\sqrt{ -3/\bar{\Lambda} }} \frac{D_a^-}{D_a^+}, \quad
		\mathcal{T}^p_l = \frac{1}{\sqrt{ -3/\bar{\Lambda} }} \frac{D_p^-}{D_p^+}.
	\end{align}
	
	Below, we use the numerical integration method to calculate TLNs.

	\subsection{Solution 1} 
	\begin{figure}[!htb]
		\centering
		\subfigure{
			\includegraphics[width=0.38\textwidth]{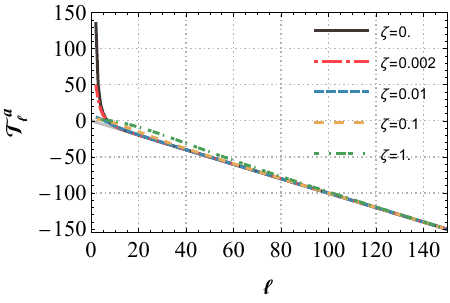}
			\label{G1_TLN_Va}
		}
		\subfigure{
			\includegraphics[width=0.38\textwidth]{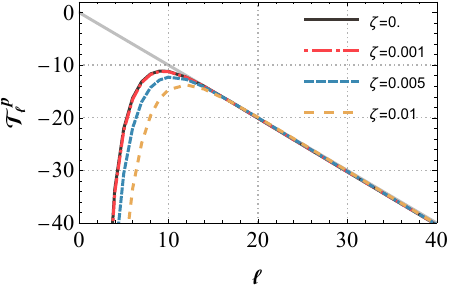}
			\label{G1_TLN_Vp}
		}
		\caption{The TLNs $\mathcal{T}^a_l$ and $\mathcal{T}^p_l$ versus $\ell$ with cosmological constant $\bar{\Lambda}= -30000$ for the axial(left) and polar(right) perturbation. }
		\label{G1_TLN_Va_Vp}
	\end{figure}
	In Fig.\ref{G1_TLN_Va_Vp}, we show the numerical results of the axial TLNs $\mathcal{T}^a_l$ and polar TLNs $\mathcal{T}^p_l$ for different values of $\zeta$ when $\bar{\Lambda}=-30000$.    
	It can be seen that  the  $\mathcal{T}^a_l$ gradually decreases and approaches $-\ell$ as $\zeta$ increases  near $\ell=2$ .   
	However, for $\ell \approx 20$, the increase of $\mathcal{T}^a_l$ will instead make the difference between $\mathcal{T}^a_l$ and $-\ell$ more obvious.
	Whatsmore, as $\zeta$ increases and $\ell$ approaches $2$, the difference between $\mathcal{T}^p_l$ and $-\ell$ becomes more and more obvious. When $\zeta$ reaches certain critical values, near $\ell = 2$, $\mathcal{T}^p_l$ may diverge. This is because there is a singularity between the event horizon and the cosmological horizon in the perturbation equation, causing a divergent term to appear in the numerical integration process.
	Note that when $\zeta=0$, the Love number result returns to the AdS case in Ref.\cite{Tidal_Franzin_2024}.

\begin{figure*}[!htb]
		\centering
		\subfigure[]{
			\includegraphics[width=0.31\textwidth]{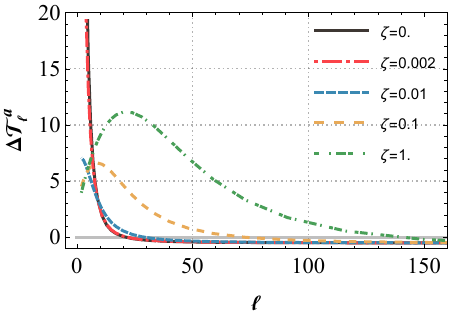}
			\label{G1_TLNl_Lambdan30000_Va}
		}
		\subfigure[]{
			\includegraphics[width=0.31\textwidth]{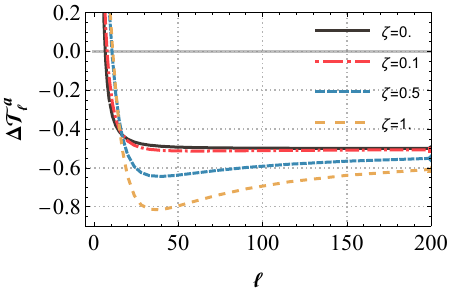}
			\label{G1_TLNl_Lambdan300_Va}
		}
		\subfigure[]{
			\includegraphics[width=0.31\textwidth]{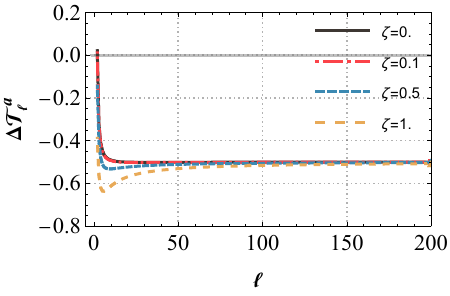}
			\label{G1_TLNl_Lambdan3_Va}
		}
		\caption{The axial $\Delta\mathcal{T}^a_l$  versus $\ell$ for (a)$\bar{\Lambda}=-30000$, (b)$\bar{\Lambda}=-300$ and (c)$\bar{\Lambda}=-3$.}
		\label{G1_TLNl_Va}
	\end{figure*}

	Fig.\ref{G1_TLNl_Lambdan30000_Va}-\ref{G1_TLNl_Lambdan3_Va} further describes the variation $\Delta\mathcal{T}^a_l=\mathcal{T}^a_l+\ell$ as a functon of $\ell$.
	For $\bar{\Lambda}=-30000$, when $\zeta$ is large enough, $\Delta\mathcal{T}^a_l$ first increases and then decreases with $\ell$, obviously.   
	However, when $\bar{\Lambda}=-300$ or $\bar{\Lambda}=-3$, these phenomena will become different, but similarly, as $\zeta$ increases, $\Delta\mathcal{T}^a_l$ will deviate from $\Delta\mathcal{T}^a_l[\zeta = 0]$ at certain values of $\ell$.
	
\begin{figure*}[!htb]
		\centering
		\subfigure[]{
			\includegraphics[width=0.31\textwidth]{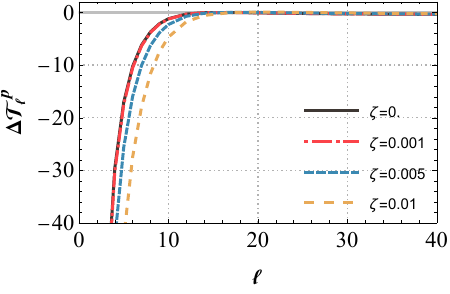}
			\label{G1_TLNl_Lambdan30000_Vp}
		}
		\subfigure[]{
			\includegraphics[width=0.31\textwidth]{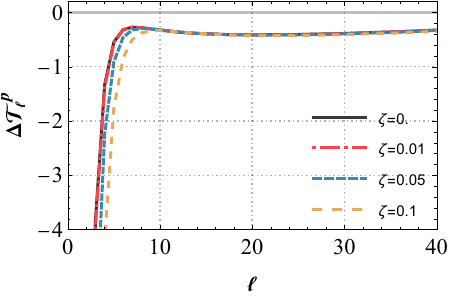}
			\label{G1_TLNl_Lambdan300_Vp}
		}
		\subfigure[]{
			\includegraphics[width=0.31\textwidth]{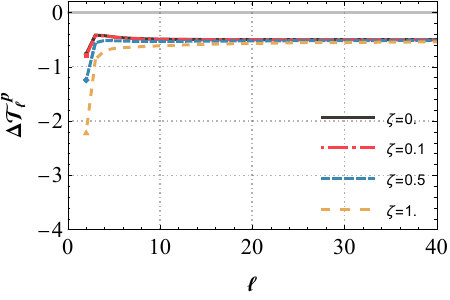}
			\label{G1_TLNl_Lambdan3_Vp}
		}
		\caption{The poral  $\Delta\mathcal{T}^p_l$ versus $\ell$ for (a)$\bar{\Lambda}=-30000$, (b)$\bar{\Lambda}=-300$ and (c)$\bar{\Lambda}=-3$.}
	\end{figure*}

	Similarly, to describe in more detail the variation $\Delta\mathcal{T}^p_l=\mathcal{T}^p_l+\ell$, we present Figs.\ref{G1_TLNl_Lambdan30000_Vp}-\ref{G1_TLNl_Lambdan3_Vp}.
	Obviously, as $\zeta = 0$ increases, $\Delta\mathcal{T}^a_l$ will decrease monotonically, and this phenomenon will become more obvious near $\ell = 2$.
	
	\subsection{Solution 2} 
	As for solution 2, during the numerical solution process, there will be several additional coefficient equations that do not exist in the solution 1 or the AdS case, which results in only the trivial solution of $0$ being obtained. This may be due to $ h[r] \neq 1$ and the ambiguity in the effective cosmological constant $\bar{\Lambda}$ defined solely by $f[r]$. 
	Therefore, this requires us to flexibly choose the effective cosmological constant $\bar{\Lambda}$ based on the condition that a non-trivial solution can be obtained during the analysis. For example, in the process of solving the axial TLNs,  we choose  $\bar{\Lambda}= \Lambda  \left(1-{\zeta^2 \Lambda }/{3}\right)$, in contrast, this assumption will fail when solving  the polar TLNs.
	
	\begin{figure}[!htb]
		\centering
		\subfigure{
			\includegraphics[width=0.38\textwidth]{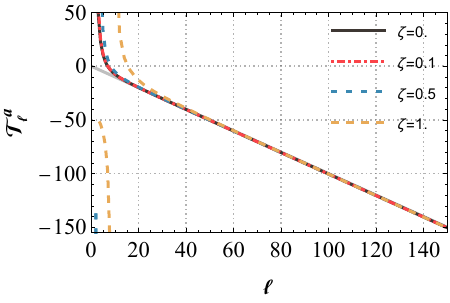}
		}
		\caption{The axial TLNs $\mathcal{T}^a_l$ versus $\ell$ with $\bar{\Lambda}= -30000$. }
		\label{G2_TLN_Va}
	\end{figure}
	In Fig.\ref{G2_TLN_Va}, we show the numerical results of the axial TLNs $\mathcal{T}^a_l$ for different values of $\zeta$ when $\bar{\Lambda}=-30000$.
	It can be seen that near $\ell=2$, as $\zeta$ increases, $\mathcal{T}^a_l$ becomes further away from $-\ell$. Until when $\zeta$ exceeds a critical value, a new branch appears near $\ell=2$. This is opposite to the result of solution 1.
	
\begin{figure*}[!htb]
		\centering
		\subfigure[]{
			\includegraphics[width=0.31\textwidth]{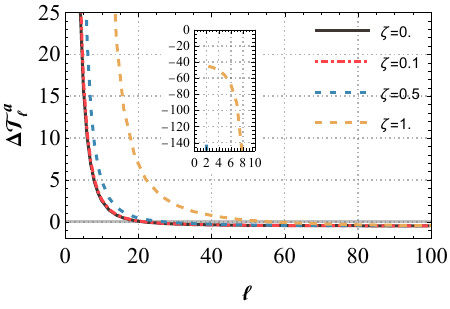}
			\label{G2_TLNl_Lambdan30000_Va}
		}
		\subfigure[]{
			\includegraphics[width=0.31\textwidth]{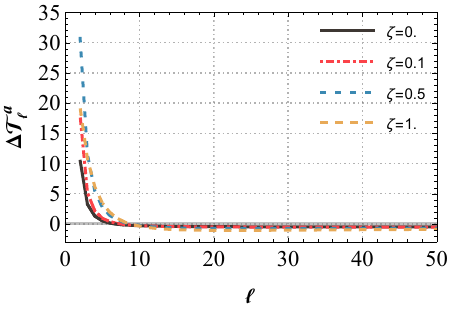}
			\label{G2_TLNl_Lambdan300_Va}
		}
		\subfigure[]{
			\includegraphics[width=0.31\textwidth]{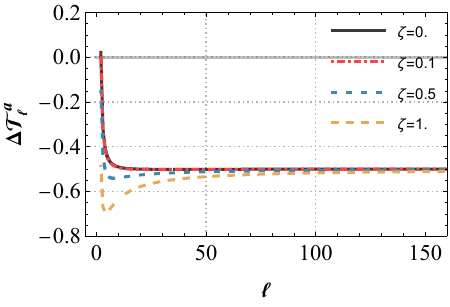}
			\label{G2_TLNl_Lambdan3_Va}
		}
		\caption{The axial $\Delta\mathcal{T}^a_l$  versus $\ell$ for (a)$\bar{\Lambda}=-30000$, (b)$\bar{\Lambda}=-300$ and (c)$\bar{\Lambda}=-3$.}
		\label{G2_TLNl_Va}
	\end{figure*}

	Similarly, Fig.\ref{G2_TLNl_Va} describes the $\Delta\mathcal{T}^a_l$ as a functon of $\ell$. 
	For $\bar{\Lambda}=-30000$, as $\zeta$ increases, during the decrease of $\ell$, the increase of $\Delta\mathcal{T}^a_l$ becomes more and more obvious until a new branch appears. A similar phenomenon can also be observed in the graph \ref{G2_TLNl_Lambdan300_Va} for $\bar{\Lambda}=-300$.
	In contrast, when $\bar{\Lambda}=-3$, it is similar to the phenomenon in Fig.\ref{G1_TLNl_Lambdan3_Va}, but the influence of $\zeta$ is more obvious in solution 2.

	\section{Conclusion}\label{conclusion}
	In this study, we explored the QNMs and TLNs of covariant quantum-modified black hole models with cosmological constant, focusing on two distinct effective metric solutions in Ref. \cite{Covariant_Yang_2025}. Through deriving axial and polar perturbation equations, we employed the pseudospectral method to calculate QNMs and numerical approaches to determine TLNs, aiming to clarify how quantum effects influence these physical quantities.
	
	For QNMs, notable differences emerged between the two solutions. In Solution 1, purely imaginary QNMs remained purely imaginary as $\zeta$ varied, maintaining their intrinsic properties. In contrast, Solution 2 exhibited more complex behavior: some purely imaginary QNMs of both polar and axial perturbations transitioned to complex numbers with changes in $\zeta$, indicating a more pronounced interplay between quantum effects and spacetime perturbations in this case. Additionally, across both solutions, the impact of $\zeta$ became increasingly significant for higher overtone modes.
	
	Regarding TLN, we have also found that the influence of $\zeta$ exhibits different patterns of the two solutions. For soluton 1,  when $\zeta$ exceeded a threshold, $\Delta\mathcal{T}^a_l$ reached a maximum at a specific angular quantum number $\ell$ before gradually decreasing. In contrast, the influence of $\zeta$ on $\Delta\mathcal{T}^p_l$ for polar perturbations was found to be monotonic. 
	For solution 2, as $\zeta$ increases, $\Delta\mathcal{T}^a_l$ also shows notable changes near $\ell=2$.
	In conclusion, the influence of $\zeta$ on the TLNs is significant when $\ell$ is small, but negligible when $\ell$ is sufficiently large.
	
	These findings highlight the sensitivity of black hole perturbation properties to the form of quantum corrections (as embodied by the two metric solutions) and the magnitude of the loop quantum parameter $\zeta$. They provide insights into how quantum effects might imprint distinct signatures in gravitational wave signals and tidal interactions, offering potential avenues for probing quantum gravity effects through future gravitational wave observations. Further investigations into quantum corrected black hole models could deepen our understanding of the interplay between quantum effects and gravity.
	
	\begin{acknowledgments}
		This work is supported by National Natural Science Foundation of China (NSFC) with Grant No. 12275087. 
	\end{acknowledgments}
	
	\appendix

	\section{Pseudospectral Method}\label{CPS}
	
	In determining QNMs using the pseudospectral method, we employ Eddington-Finkelstein coordinates, in which imposing boundary conditions is more straightforward \cite{Revisiting_Mamani_2022}.
	
	First, defining the parameter $u[r_*]=1 /r[r_*]$, we obtain:  
	\begin{align}
		\frac{du[r_*]}{dr_*}= \frac{d}{dr_*} \left(\frac{1}{r[r_*]}\right) =-\frac{ f[r_*] \sqrt{h[r_*]}}{r[r_*]^2}.
	\end{align}  
	To transform the Eq.\eqref{MainEQ} to Eddington-Finkelstein coordinates, we must use the following transformation:
	\begin{align}
		\Phi_s[u]= \frac{\phi_s[u]}{u} e^{- i \omega r_*[u]}.
	\end{align} 
	The equation can be rewritten as:  
	\begin{widetext}
	\begin{align}
		0  =& \phi_s[u] \left(-\frac{  V\left[1/u,s\right]}{u f\left[1/u\right]}+ h[1/u] f'[1/u]+\frac{1}{2} f[1/u] h'[1/u]-2 i \omega  \sqrt{h[1/u]}\right) \notag\\
		&+\phi_s'[u] \left(- u h[1/u] f'[1/u]-\frac{1}{2}  u f[1/u] h'[1/u]+2 i  \omega  u \sqrt{h[1/u]}\right)  + \phi_s''[u] u^3 f[1/u] h[1/u].
	\end{align}
	\end{widetext} 
	Here, this result recovers the dS case and aligns with results in the Ref.\cite{Overdamped_Jansen_2017}.  
	At infinity, the metric is asymptotically de Sitter (dS) spacetime.  
	Thus, we can assume
	\begin{align}
		\phi_s[u] =  \frac{1}{u-u_c} (u-u_c)^{-{2 i \omega }/{\kappa_c}}   y_s[u],
	\end{align} 
	where $\kappa_c$ is the surface gravity at the cosmological horizon, satisfying $\kappa_c=-1/2 f'[1/u_c]$.
	 
	On the other hand, to map the domain from $[1/u_h,1/u_c]$ to $[0,1]$, we perform the transformation $u = u_c + z (u_h-u_c)$.
	Note that, when $z=1$, $u=u_h$, and when $z=0$, $u=u_c$.   
	Finally, the perturbation equation can be transformed to:  
	\begin{align}
		0 =A_0 y_s[z]  + A_1 y'_s[z] + A_2 y''_s[z],
	\end{align}
	where 
	\begin{widetext}
		\begin{align}
			A_0=
			&
			\omega^2 \left(-8 f\left[{1}/{u}\right] h\left[{1}/{u}\right] u^4 + 8 z \sqrt{h\left[{1}/{u}\right]} u^2 \left(u_c - u_h\right) f'\left[{1}/{u_c}\right]\right) \notag\\
			&- 2 i \omega u f'\left[{1}/{u_c}\right] \bigg(2 z \sqrt{h\left[{1}/{u}\right]} \left(u_c - u_h\right) \big(\sqrt{h\left[{1}/{u}\right]} \left((-1 + z) u_c - z u_h\right) f'\left[{1}/{u}\right] \notag\\
			&+ \left((-1 + 2 z) u_c - 2 z u_h\right) f'\left[{1}/{u_c}\right]\big) + f\left[{1}/{u}\right] u \left(6 h\left[{1}/{u}\right] u^2 + z \left(u_h-u_c \right) h'\left[{1}/{u}\right]\right)\bigg) \notag\\
			&+ f'\left[{1}/{u_c}\right]^2   \bigg(-2 z^2 \left(u_c - u_h\right)^2 \frac{V_{\text{eff}}\left[{1}/{u}, s\right]}{f\left[{1}/{u}\right]} + u \big(2 z h\left[{1}/{u}\right] \left(u_c - u_h\right) \left((-1 + 2 z) u_c - 2 z u_h\right) f'\left[{1}/{u}\right]  \notag\\
			& + f\left[{1}/{u}\right] \left(4 h\left[{1}/{u}\right] u^3 + z \left(u_c - u_h\right) \left((-1 + 2 z) u_c - 2 z u_h\right) h'\left[{1}/{u}\right]\right)\big)\bigg),
			\\
			A_1&=
			\omega \left(8 i z f\left[{1}/{u}\right] h\left[{1}/{u}\right] u^4 f'\left[{1}/{u_c}\right] - 4 i z^2 \sqrt{h\left[{1}/{u}\right]} u^2 \left(u_c - u_h\right) f'\left[{1}/{u_c}\right]^2\right) \notag\\ 
			&- z u^2 f'\left[{1}/{u_c}\right]^2 \left(2 z h\left[{1}/{u}\right] \left(u_h-u_c \right) f'\left[{1}/{u}\right] + f\left[{1}/{u}\right] \left(4 h\left[{1}/{u}\right] u^2 + z \left(u_h-u_c \right) h'\left[{1}/{u}\right]\right)\right),\\
			A_2&=
			2 z^2 f\left[{1}/{u}\right] h\left[{1}/{u}\right] u^4 f'\left[{1}/{u_c}\right]^2.
		\end{align}
	\end{widetext} 
	where the $V_{eff}$  can be replaced by $V_s$, $V_a$ and $V_p$.
	
	\section{Expansion coefficients} \label{CTLN}
	To calculate the expansion coefficients, first, we substitute Eqs.\eqref{PhiaInf}-\eqref{Phiprh} into Eqs.\eqref{PloEq}-\eqref{AxiEq} with $\omega=0$, respectively. Then, by comparing the coefficients of each order, we obtain a series of equations. Solving these equations yields the expressions for the coefficients of each order.
	Note that, in the process of calculating the coefficients, there are always some undetermined degrees of freedom, such as $\{a^+_0,a^-_0,p^+_0,p^-_0,p^h_0,a^h_0\}$. Here, we always choose $a^+_0=a^-_0=p^+_0=p^-_0=p^h_0=a^h_0 = 1$ for normalization. All other undetermined free coefficients are chosen to be zero.
	
	\subsection{Solution 1}
	The coefficients of asymptotic expansion at infinity for the axial perturbation are 
	\begin{widetext}
	\begin{align}
		a^+_1 &= 0,  									& a^-_1 &= 0,\\
		a^+_2 &= -\frac{3 \left(6 + 3 m - 4 \zeta^2 \Lambda\right)}{2 \Lambda \left(3 - \zeta^2 \Lambda\right)}, &a^-_2 &= -\frac{3 m}{2 \Lambda \left(3 - \zeta^2 \Lambda\right)}, \\
		a^+_3 &= \frac{3 \left(3 - 2 \zeta^2 \Lambda\right)}{\Lambda \left(3 - \zeta^2 \Lambda\right)},  &a^-_3 &= 0,\\ 
		a^+_4 &= \frac{27 (-4 + m) (2 + m) + 36 \zeta^2 (5 + m) \Lambda - 60 \zeta^4 \Lambda^2}{8 \Lambda^2 \left(3 - \zeta^2 \Lambda\right)^2}, &a^-_4 &=\frac{9 \left(3 (-10 + m) m + 4 \zeta^2 (-3 + 5 m) \Lambda + 4 \zeta^4 \Lambda^2\right)}{40 \Lambda^2 \left(3 - \zeta^2 \Lambda\right)^2},
	\end{align}
	\end{widetext} 
	Similarly, for the polar perturbation, the coefficients of asymptotic expansion of $\Phi_{p}$ at infinity are 
	\begin{widetext}
	\begin{align}
		p^+_1 &= 0 ,  	\quad \quad \quad p^-_1 = 0,\\
		p^+_2 &= \frac{3 \zeta^2}{m} + \frac{4 \left(3 - 2 \zeta^2 \Lambda\right)^2}{m^2} - \frac{3 \left(6 + 3 m - 4 \zeta^2 \Lambda\right)}{2 \Lambda \left(3 - \zeta^2 \Lambda\right)},  	\quad
		p^-_2 = \frac{\zeta^2}{m} + \frac{4  \left(3 - 2 \zeta^2 \Lambda\right)^2}{3 m^2} + \frac{3 m}{2 \Lambda \left(-3 + \zeta^2 \Lambda\right)}, \\
		p^+_3 &= \frac{3}{\Lambda} \bigg( \frac{3 - 2 \zeta^2 \Lambda}{3 - \zeta^2 \Lambda} - \frac{16 \Lambda \left(3 - 2 \zeta^2 \Lambda\right)^3}{9 m^3}  
		+ \frac{ 4 \left(9 - 21 \zeta^2 \Lambda + 7 \zeta^4 \Lambda^2\right)}{3 m \left(3 - \zeta^2 \Lambda\right)} 
		- \frac{12 \zeta^2 \Lambda \left(9 - 9 \zeta^2 \Lambda + 2 \zeta^4 \Lambda^2\right)}{3 m^2 \left(3 - \zeta^2 \Lambda\right)}\bigg)
		,  \notag\\
		p^-_3 &= \frac{6}{m \Lambda} \bigg( -\frac{7 \zeta^2 \Lambda}{3} + \frac{3}{3 - \zeta^2 \Lambda} - \frac{\zeta^2 \Lambda \left(3 - 2 \zeta^2 \Lambda\right)}{m} 
	- \frac{4  \Lambda \left(3 - 2 \zeta^2 \Lambda\right)^3}{9 m^2} \bigg).	
	\end{align}
	\end{widetext} 
	The coefficients of asymptotic expansion at the horizon are  
	\begin{widetext}
		\begin{align}
			a^h_1 &= \frac{-6 + (2 + m) r_h}{2 (3 - r_h)} + \frac{\zeta^2 \left(24  - 12 r_h + r_h^2\right)}{(3 - r_h) r_h^3},  \\
			a^h_2 &= 
			\frac{144  - 24 (4 + m) r_h + (2 + m) (8 + m) r_h^2}{16 (-3  + r_h)^2} - \frac{\zeta^2 \left(576 - 6 (78 + m) r_h + 114  r_h^2 + (-7 + m) r_h^3\right)}{4 r_h^3 (-3  + r_h)^2} \notag\\
			& - \frac{3 \zeta^4 (-2  + r_h)^2 \left(24 - 12 r_h + r_h^2\right)}{4 r_h^6 (-3 + r_h)^2}
			, \\
			p^h_1 &= \frac{1}{a_{hf}}\bigg(-r_h^6 \left(36 - 12  r_h + m (2 + m) r_h^2\right) + 4 \zeta^2 r_h^3 \left(108  + 6 (-14 + 3 m)  r_h - 12 (-2 + m) r_h^2 + (-3 + 2 m) r_h^3\right) \notag\\
			&- 12 \zeta^4 (-2  + r_h)^2 \left(24 - 12 r_h + r_h^2\right) \bigg) , \\
			p_{hf} &=
			2 (3 - r_h) r_h^3 \left(6 \zeta^2 (r_h-2)^2 - r_h^3 (6 + m r_h)\right).
		\end{align}
	\end{widetext} 
	The expression for $p^h_2$ is complex and thus not shown here.
	Meanwhile, when $\zeta=0$, the result recovers that in reference \cite{Tidal_Franzin_2024}.
	
	\subsection{Solution 2}
	Similarly,   by substituting Eqs.\eqref{PhiaInf}-\eqref{Phiprh} into Eqs.\eqref{PloEq}-\eqref{AxiEq}, a series of equations can be obtained. Through calculation, we can get the coefficients of asymptotic expansion at infinity as
	\begin{widetext}
		\begin{align}
			a^+_1 &= 0,  									& a^-_1 &= 0,\\
			a^+_2 &= - \frac{3 \left(6+3 m-5 \zeta^2 \Lambda\right)}{2 \Lambda  \left(3 - \zeta^2 \Lambda \right)}, &a^-_2 &= -\frac{3 m-2 \zeta^2 \Lambda }{6 \Lambda -2 \zeta^2 \Lambda^2}, \\
			a^+_3 &= \frac{3}{\Lambda }-\frac{3 \zeta^2}{6-2 \zeta^2 \Lambda },  &a^-_3 &= 0,\\ 
			a^+_4 &= \frac{27\left(-4 + m\right)\left(2 + m\right) + 252\zeta^{2}\Lambda - 69\zeta^{4}\Lambda^{2}}{8\Lambda^{2}\left(-3 + \zeta^{2}\Lambda\right)^{2}}, &a^-_4 &= \frac{3\left(9\left(-10 + m\right)m + 6\zeta^{2}\left(16 + 7m\right)\Lambda - 44\zeta^{4}\Lambda^{2}\right)}{40\Lambda^{2}\left(-3 + \zeta^{2}\Lambda\right)^{2}},
		\end{align}
	\end{widetext} 
	
	Finally, the coefficients of asymptotic expansion at the horizon are  
	\begin{widetext}
		\begin{align}
			a^h_1 &= 
			\frac{-6 + (2 + m) r_h}{2 \left(3 - r_h\right)} - \frac{\zeta^2 \left(3 - 6 r_h + 2 r_h^2\right)}{\left(3 - r_h\right) r_h^3}
			,  \\
			a^h_2 &= \frac{144 - 24 (4 + m) r_h + (2 + m) (8 + m) r_h^2}{16 \left(-3 + r_h\right)^2} - \frac{\zeta^2 \left(-144 + 3 (90 + 11 m) r_h - 6 (24 + 5 m) r_h^2 + (24 + 7 m) r_h^3\right)}{8 \left(-3 + r_h\right)^2 r_h^3}  \notag\\
			&+
			\frac{\zeta^4 \left(3 - 6 r_h + 2 r_h^2\right) \left(30 - 24 r_h + 5 r_h^2\right)}{4 \left(-3 + r_h\right)^2 r_h^6}.
		\end{align}
	\end{widetext}
	When $\zeta=0$, the result recovers that in reference \cite{Tidal_Franzin_2024}.
	
	\bibliographystyle{unsrt}

\end{document}